\documentstyle[preprint,prb,aps]{revtex}
\begin{document}
\draft
\title{Possibility of Kauzmann points in the vortex matter phase diagram of single
crystal YBa$_2$Cu$_3$O$_{7-\delta}$}
\author{S. B. Roy$^{1,2,+}$, Y. Radzyner$^{1}$, D. Giller$^{1}$, Y. Wolfus$^{1}$, A.
Shaulov$^{1}$, P. Chaddah$^{2}$ and Y. Yeshurun$^{1}$}
\address{$^{1}$Department of Physics, Institute of Superconductivity, Bar-Ilan\\
University, 52900 Ramat-Gan, Israel}
\address{$^{2}$Low Temperature Physics Laboratory, Centre for Advanced Technology,\\
Indore 452013, India}
\date{\today}
\maketitle

\begin{abstract}
We highlight interesting thermomagnetic
history effects across the transition line between the (quasi) ordered and
disordered vortex states in single crystal YBa$_2$Cu$_3$O$_{7-\delta}$, and
argue that these features are indicative of the first order nature of
the transition line. We suggest that the destruction of the ordered vortex
state in YBa$_2$Cu$_3$O$_{7-\delta}$ leading to vortex liquid (at high
temperatures and low fields) and amorphous vortex solid (at low 
temperatures and high fields), takes place along a unified first-order 
transition line. The nonmonotonic behavior of this first order transition line gives rise to
the possibility of more than one Kauzmann point where the entropies of the ordered and
disordered vortex states are equal. In the high temperature region, one may order the
vortex lattice by warming it, giving rise to an inverse melting effect.
\end{abstract}

In a recent report Avraham et al \cite{1} has shown that the destruction
of the quasi-ordered vortex lattice or Bragg glass \cite{2} in single
crystal samples of Bi$_2$Sr$_2$CaCu$_2$O$_8$ (BSCCO) takes place along a
unified first order transition line. Two different types of energy -- thermal energy and pinning energy --
actually compete with the elastic energy leading to the destruction of
the ordered vortex lattice. At low temperatures and high fields pinning
dominates, leading to a field/disorder induced destruction of the ordered
vortex lattice. At high temperatures this unified transition line changes its
character from disorder induced transition to thermally-induced melting.
The apparently unusual finding is the non-monotonic nature of this
first-order transition line, leading to the paradoxical situation
that in a certain field-temperature (B-T) window the ordered vortex state
has larger entropy than the disordered vortex state. This in turn has the
interesting implication that a crystal transforms into liquid or amorphous
state on decreasing the temperature. Such a situation is quite rare but
not unknown, one classic example being the the melting curve
of $^3$He showing a pressure minimum \cite{3}. Similar situation also exists
in spin lattice systems\cite{4} where the spin-glass state transforms into a
long-range magnetic ordered state with the increase in temperature.
However, such a transition is known to be a continuous transition with 
definite critical response\cite{4}.

The same non-monotonic character of the transition line between two kinds of
vortex solids has been reported earlier for YBa$_{2}$Cu$_{3}$O$_{7-\delta }$ (YBCO) single
crystals \cite{5,6}. In addition the presence of metastability has also been
highlighted across this phase transition \cite{6,7}. We extend this study to
argue that the vortex solid-solid transition line in YBCO is also
a first order transition line, and there exists a situation of "ordering
by heating'' in a certain B-T window of YBCO as well. We shall use the
minor hysteresis loop (MHL) technique \cite{8} to study the first order
transition line. This technique, although less rigorous than an actual
equilibrium thermodynamic measurement, has been quite successful recently in
studying the first-order nature of the vortex solid-solid transition in low
tempearture superconductors like CeRu$_2$ \cite{9,10,11} and
NbSe$_{2}$ \cite{12,13}, and in high temperature superconductors such as
YBCO \cite{6} and LaSrCuO\cite{14}.

Local magnetization measurements using an array of
10 $\times$ 10$\mu$m$^2$ Hall sensors (sensitivity better than 0.1 G) were
carried out on a 0.5$\times$0.3$\times$0.02 mm$^3$ untwinned YBaCuO
single crystal \cite{15}.

A typical field dependence of magnetization showing peak-effect,
which we use to track the vortex solid-solid phase transition line, 
is shown in Fig. 1. We identify four characteristic fields:
B$^+_{onset}$, B$^-_{onset}$, B$^+_{kink}$ and B$^-_{kink}$,
marked by arrows in Fig.1. We have earlier argued that the solid-solid
thermodynamic transition field B$_{SS}$ can actually be identified with the B$^-_{kink}$ 
\cite{6}. Collating these characteristic field B$^-_{kink}$ at various 
temperatures from our isothermal magnetization studies we present a 
B-T phase diagram in Fig.2 showing this B$^-_{kink}$(T) line.
This is similar to the (B-T) phase diagram that has been reported
earlier\cite{6}, but is reproduced here again to make the present study a 
self contained one.

It is apparent from Fig. 2 that the non-monotonic
nature of the phase transition line is more prominent in comparison to
BSCCO \cite{1}. The slope of the transition line changes sign twice as a
function of temperature, first at around 50K going from negative to positive,
and then at around 75K back to negative again. We shall now concentrate on
the 50K-75K regime of this phase transition line where,
akin to that in BSCCO \cite{1}, exists the interesting
suggestion of the transition from the disordered-solid to ordered solid
achieved by heating. We present results in the form of MHLs obtained after
preparing the vortex state following two distinct experimental protocols:
\begin{enumerate}
\item  Zero field cool (ZFC) the sample to the temperature of measurement and
then increase the field to go to the vortex solid-solid phase transition region
denoted by the shaded area in Fig.2. The field is then lowered to zero so
that an MHL is obtained.
\item  Cross the vortex solid-solid phase transition line by varying
temperature in the presence of an external field. To do this in the 50K-75K regime,
which had not been explored before, we
cool the sample from above T$_{C}$ to 50K in the presence of external field
of 40 kOe. We then lower the field to the target value at 50K,
and then increase the temperature to the temperature of measurement.
This is the counterpart of the step-down procedure used in Ref.6, where only
the low-temperature region was explored.
\end{enumerate}

In Fig. 3 we present MHLs obtained under the protocol 1 at 70K. These MHLs
are obtained by terminating the field increasing cycle of the M-B curve
at various points in the field regime B$^-_{onset}\leq$ B$\leq B^+_{kink}$.
Within the Bean critical state model (which explains well the irreversible
magnetization of a type-II superconductor) such MHLs are expected to meet the enevlope
M-B curve after the field is decreased by an amount 2B$^*$, where B$^*$ is
the field for full penetration\cite{16} at that particular B. In fact the MHLs
drawn at various fields B $>B^+_{kink}$ show this expected
behaviour. However, the behaviour changes markedly at the onset of the peak
effect (PE) regime, and the MHLs obtained by procedure 1 (ZFC) saturate 
without reaching the the upper envelope curve (see
Fig. 3). Since the amount of hysteresis $\Delta $M is proportional to the
size D of the sample exhibiting pinning, it is clear that the vortex disordered
solid (with enhnaced pinning) starts nucleating at B $>B^+_{onset}$ and its
formation is complete only at $B^+_{kink}$ where the MHLs saturate only on
reaching the upper envelope curve. This kind of nucleation and growth of the
enhanced pinning state are typical characteristics of a first order
transition, and have earlier been observed in low temperature 
superconductors like CeRu$_{2}$ \cite{8,9,10,11} and NbSe$_{2}$ \cite{12,13}, YBCO\cite{6} and
LaSrCuO\cite{14}.

In Fig.4 we present MHLs at 70K obtained under
the experimental protocol 2. The MHLs obtained both by increasing and
decreasing B overshoot the enevelope curve. This in turn implies
that flux pinning obtained in this manner is more than that obtained in
crossing the vortex solid-solid transition line by isothermal field
variation. It has been argued earlier that extent of metastability (i.e.
supercooling/superheating) associated with a first order transition is more
if the transition line is crossed by the variation of temperature in presence
of a constant magnetic field in comparison to the situation where the line
is crossed by isothermal variation of the applied field \cite{17}. This is because the
variation of the magnetic field produces fluctuation which drives the metastable state
in the local minimum of the free energy curve to
the stable state across the energy barrier. We argue that the observed
overshooting of the MHLs in Fig. 4 is thus another indication of the first order nature of
the vortex solid-solid transition line.

It is to be noted here that the metastability
across the first order vortex solid-solid transition line has been
highlighted recently in BSCCO \cite{1,18,19}. In fact the associated irreversibility in
this transition region was removed by using an applied
ac field H$_{ac}$ to reveal the step-jump in the magnetization, which
in turn was used to establish the first order nature of the transition line 
\cite{1}. In contrast we use the metastable characteristic across the
transition line itself to identify its first order nature. We have earlier
used the same technique to study the nature of the vortex 
solid-solid transition line of YBCO below 50K (Ref. 6).

Combining the present study as well as the earlier ones on 
YBCO \cite{5,6} we argue that as in BSCCO the extended B$_{SS}$(T) line 
coincides with the melting line vortex ordered solid B$_M$(T) 
line at high temperatures. Hence the
disordering of the vortex ordered solid is apparently always a first order
transition. We note that B$_{SS}$(T) being a phase transition line allows one
to assert that the free energies of the two solid phases are equal along
this line, and they satisfy the inequality of opposite signs as the B$_{SS}$
(T) line is crossed. Our conclusion that this line corresponds to a first order
phase transition implies, in addition, that the entropies of the two
solid phases viz. the 'ordered' Bragg glass and the 'disordered' vortex
glass, are unequal along this line.
The shape of this combined transition line, however, gives rise
to many interesting possibilities, including two Kauzmann points \cite{20}.
In the present (B-T) phase diagram of YBCO (Fig. 2) we define these Kauzmann
points as the points where the slope of B$_{SS}$ changes its sign - once
around 50K and then at around 75K.
At a Kauzmann point the entropies of the disordered
and ordered state are equal. The Kauzmann point is well known in the context
of (molecular) liquid-glass transition, where to avoid entropy crisis below T$
_{Kauzmann}$ the liquid is frozen into a glassy state at a 
temperature T$_{G}>$T$_{Kauzzman}$.
On the other hand, a Kauzmann point can actually be reached in the
pressure(P)-temperature(T) phase diagram of $^3$He. It leads to the
apparently anomalous situation where the solid entropy is higher than
the liquid entropy \cite{3}. This is of course now understood in terms of the
(nuclear) spin contribution to entropy which due to Pauli exclusion
principle is less for liquid than for the solid. At very low temperatures the spin contribution
dominates over the structural contributions, and thus the entropy per atom
of the solid is greater than the liquid. Experiments also indicate that
poly(4-methylpentene-1) exhibits a pressure maximum, hence a Kauzmann
point along its melting curve (see Ref. 20 and references therein).
The interpretation of this observation becomes complicated because of the
appearence of an additional phase (Ref.20). In the proposed (B-T) phase
diagram of oxide superconductors with reentrant melting of a vortex solid 
\cite{21}, there would be a Kauzmann point at the 'tip of the nose'.
This phase diagram, however, still has to be confirmed experimentally.
With these information we now attempt to rationalize the entropic 
relations between various phases in the vortex matter phase diagram of YBCO.

Between two Kauzmann points (50K,($\approx$)1.1T and 72K,($\approx$)2.1T)
in the B-T phase diagram of YBCO, the dB$_{SS}$/dT
has a positive slope; and from the Clausius-Clapeyron relation this will
indicate a neagative entropy change across this first order transition line. (The
implicit assumption here is that the equilibrium magnetization shows a
positive jump at the phase transition point as in BSCCO). A negative entropy leads to
the paradoxical situation where the ordered vortex solid has larger entropy
than the disordered vortex solid. The name ordered vortex solid itself
indicates that it is structurally more ordered. Hence, as in the case of solid $^3$He
the extra entropy needs to be attributed to some additional degrees of freedom.
In the same vein as in BSCCO (Ref. 1) it can be argued that in the disordered
vortex solid the the flux-lines or vortices wander out of their unit cells
and become entangled. However, their fluctuations are small on short time scales.
In contrast, in the ordered vortex solid there is no large scale wandering of
the vortex and entanglement. But the effect of thermal fluctuations within
the unit cell is comparatively large, which in turn results in larger entropy.

An important aspect which is not commonly addressed in the studies of vortex
matter is the electronic structure of the vortex cores. The existence of bound
quasi particle states in the normal vortex cores of a 
conventional superconductor has been predicted
in early sixties by Caroli et al \cite{22} but was not established
experimentally until late eighties when Hess et al \cite{23} observed tunneling spectra in the
vortex core of NbSe$_2$ consistent with localized quasi particle states. With the
arrival of HTSC the questions are now asked regarding the low energy physics
associated with a vortex core, nodal structure (associated with the
proposed d$_{x^2 - y^2}$ wave pairings) and quasiparticle transfer between
vortices, which will ultimately govern the physical properties of the vortex
states. While localized quasi particle states inside the vortex core has now
been observed in YBCO (Ref. 24), an intersting gap like structure at the
Fermi level is found at the centre of the cores of BSCCO which scales with the
superconducting gap \cite{25}. Then there exists the theoretical prediction
that the vortex cores are magnetically ordered \cite{26}. Recent neutron
experiments on La$_{2-x}$Sr$_x$CuO$_4$ suggest that while at optimal doping individual
vortices are associated with enhnaced low frequency antiferromagnetic fluctuations,
the vortex state acquire static long-ranged antiferromagnetism in the underdoped
sample \cite{27}. In a very recent neutron experiment commensurate
antiferromagnetic ordering has also been observed in YBa$_2$Cu$_3$O$_{6.5}$
(T$_{C}$=55K) (Ref. 28).

With these informations on the quasiparticles in the vortex cores of the HTSC
materials, one can probably bring the analogy between the P-T diagram of 
$^3$He and vortex matter phase diagram a bit closer. In an ordered vortex solid
the quasi particles in the individual vortex core will act independently as
the spins associated with individual $^3$He do in $^3$He solid. In the
entangled disordered vortex state the qusiparticles are likely to be more correlated,
and hence their contribution to the entropy is reduced.

We conclude that the vortex ordered solid-disordered solid
transition line in YBCO is probably a first order transition line. Combing
with the earlier results on high temperature low field melting line we suggest
that the destruction of the ordered vortex
state in YBCO takes place along a unified first-order transition line. The
non-monotonic nature of this transition line suggests the existence of two
Kauzmann points at around 75K and 50K. To explain the entropy crisis between
this two Kauzmann points, sources for additional degrees of freedom are
needed to be identified. Competition between elastic energy and pinning 
energy in the temperature region of interest and its consequence on the 
thermal contribution of entropy may be one such source. The other 
possibility related to the correlation between
quasiparticles in the individual vortex cores is also discussed.
From the experimental studies on both YBCO and BSSCO there is no indication
as yet of the vortex solid-solid transition line ending in to a critical
point, and in order to avoid further entropy crisis the slope of
this first order transition line below 50K needs to be finite and reaching
the zero value at T=0K only. In $^3$He phase diagram the T=0 point is
a Kauzmann point but without
any entropy crisis since both the ordered and the disordered vortex phase
have zero entropy at T=0 K. This need not necessarily be the case in the vortex
matter of YBCO and BSSCO where in contrast with the $^{3}$He the disordered
phase has higher entropy at low temperatures, and this disordered phase will
be relatively more susceptible to zero-point vibration at T=0 K.

ACKNOWLEDGEMENTS:

SBR acknowledges hospitality at Bar-Ilan University where the present
experimental work has been carried out. YR acknowledges support from Mifal
Hapayis - Michael Landau Foundation. YY acknowledges support from US-Israel
Binational Science Foundation. This research was supported by The Israel
Science Foundation - Centre of Excellence Programme, and by the Heinrich
Hertz Minerva Center for High Temperature Superconductivity.

+ Present and permanent adress : Low Temperature Physics Laboratory, Centre
for Advanced Technology, Indore 452013, India.

\begin{figure}[tbp]
\caption{Magnetization loop with peak effect for the YBa$_{2}$Cu$_{3}$O$_{7-%
\protect\delta }$ crystal at T=55K. Four characteristic fields B$%
_{onset}^{+} $, B$_{onset}^{-}$, B$_{kink}^{+}$ and B$_{kink}^{-}$ are
identified and marked in the figure.}
\end{figure}

\begin{figure}[tbp]
\caption{Magnetic phase diagram for YBa$_2$Cu$_3$O$_{7- \protect\delta}$
crystal showing B$^+_{onset}$(square), B$^-_{onset}$(circle), B$^+_{kink}$
(triangle), B$^-_{kink}$(diamond) lines. As discussed in Ref.6 B$^-_{kink}$%
(T) line is the vortex solid-solid phase transition line B$_{SS}$(T). Also
marked are the Kauzmann points (stars). Path marks the second field-cooling
experimental protocol (see text for details).}
\end{figure}

\begin{figure}[tbp]
\caption{Magnetization loop ( represented by diamonds) and zero field cooled
minor hysteresis loops (MHLs) following the experimental protocol 1 (see
text for details) at 70K. The MHLs were initiated by reversing the field
cycling at 10 kOe (down triangles), 12 kOe (up triangles) and 15 kOe
(circles). }
\end{figure}

\begin{figure}[tbp]
\caption{(a) Magnetization loop (represented by the solid line) and field
cooled minor hysteresis loops (MHLs) following the experimental protocol 2
(see text for details) at 70 (K). The MHLs are obtained at various target
fields both by lowering and raising field. (b) The overshooting of the FC
MHLs beyond the complete magnetization loop (solid line) is highlighted
while raising field at 8 kOe (open triangle) and 12 kOe(circle), and while
lowering the field at 8 kOe (square) and 6 kOe(circle).}
\end{figure}

\end{document}